\documentclass[12pt,a4paper]{article}
\usepackage[utf8]{inputenc}
\usepackage{amsmath,graphicx,dcolumn,subcaption,mathtools,amssymb}
\usepackage{braket}
\usepackage{siunitx}
\usepackage{appendix}
\usepackage[mathscr]{eucal}
\usepackage{bm}
\usepackage{bbm}
\usepackage{cancel}

\usepackage{color}
\usepackage{rotating}
\usepackage{xcolor}

\definecolor{darkgreen}{RGB}{0,100,0}

\usepackage[numbers,sort&compress]{natbib}
\usepackage[]{float}
\usepackage[font=small, labelfont=bf]{caption}

\usepackage{slantsc}
\usepackage{xspace}

\usepackage{setspace}
\allowdisplaybreaks

\newcommand\asbare{\alpha_{\mathrm{S}}^u} 
\newcommand\as{\alpha_{\mathrm{S}}} 
 
\newcommand\eps{\epsilon}

\def\beq{\begin{equation}} 
\def\eeq{\end{equation}} 
\def\beeq{\begin{eqnarray}} 
\def\eeeq{\end{eqnarray}}

\newcommand{\ktness}{k_T^{\text{ness}}}

\newcommand\ep{\epsilon}

\usepackage{hyperref}
\usepackage{footnotebackref}

\newcommand{\qfcut}{\mathfrak{q}_{\mathrm{cut}}}
\newcommand{\J}{\mathcal{J}}

\newcommand{\Af}{\mathfrak{A}}

\newcommand{\calPhat}{\hat{\mathcal{P}} }
\newcommand{\qf}{\mathfrak{q}}
\newcommand{\qft}{\tilde{\mathfrak{q}}}

\begin{document}
\hypersetup{pageanchor=false}
\begin{titlepage}
\begin{flushright}
  ZU-TH 54/25\\
  CERN-TH-2025-167\\
\end{flushright}

\renewcommand{\thefootnote}{\fnsymbol{footnote}}
\vspace*{0.5cm}

\begin{center}
  {\Large \bf The Quark Jet Function\\[0.3cm] for $k_T$-like Variables in NNLO QCD}
\end{center}

\par \vspace{2mm}
\begin{center}
  {\bf Luca Buonocore}$^{(a)}$, {\bf Massimiliano Grazzini}$^{(b)}$, {\bf Flavio Guadagni}$^{(b)}$\\[0.2cm]
  {\bf J\"urg Haag}$^{(c)}$ and {\bf Luca Rottoli}$^{(d)}$

\vspace{5mm}

$^{(a)}$ CERN, Theoretical Physics Department, CH-1211 Geneva 23, Switzerland\\[0.2cm]
$^{(b)}$ Physik Institut, Universit\"at Z\"urich, CH-8057 Z\"urich, Switzerland\\[0.2cm]
$^{(c)}$ Albert Einstein Center for Fundamental Physics, Institut f\"ur Theoretische Physik, Universit\"at
Bern, Sidlerstrasse 5, CH-3012 Bern, Switzerland\\[0.2cm]
$^{(d)}$ Dipartimento di Fisica G. Occhialini, Universit\`a degli Studi di Milano-Bicocca
and INFN, Sezione di Milano-Bicocca, Piazza della Scienza 3,20126 Milano, Italy\\[0.2cm]
\vspace{5mm}

\end{center}

\par \vspace{2mm}
\begin{center} {\large \bf Abstract} 

\end{center}
\begin{quote}
  \pretolerance 10000

The precise description of jet processes requires observables capable of efficiently capturing the dynamics of the energy flow
in hadronic final states.
We consider a class of tranverse-momentum like resolution variables that smoothly describe the $n+1$ to $n$ jet transition in multi-jet processes.
We discuss a general method for the computation of the corresponding quark jet function at next-to-next-to-leading order in perturbative QCD.
Rapidity divergences are regulated by using a time-like auxiliary vector.
We present explicit results for a variant of $y_{23}$ in the $E$-scheme and in the WTA scheme.

\end{quote}

\vspace*{\fill}
\begin{flushleft}
August 2025
\end{flushleft}
\end{titlepage}

\clearpage
\pagenumbering{arabic}
\setcounter{page}{1}
\hypersetup{pageanchor=true}


\renewcommand{\thefootnote}{\fnsymbol{footnote}}

\section{Introduction}
\label{sec:intro}

Jet production at high-energy colliders is one of the most fundamental probes of Quantum Chromodynamics (QCD).
At the Large Hadron Collider (LHC) jet studies are a staple in the precision physics programme. 
The vast amount of data collected so far provides important constraints in the extraction of parton distribution functions (PDFs) and in the determination of the strong coupling constant $\as$. This will be even more the case in the upcoming High-Luminosity phase of the LHC. Looking ahead, the accurate description of jet processes will be of high relevance at future $e^+e^-$ colliders.

At hadron colliders, $2 \rightarrow 2$ parton scattering constitutes the simplest jet production process.
Next-to-next-to-leading order (NNLO) QCD corrections for single-jet inclusive and dijet observables have been obtained only recently~\cite{Currie:2016bfm,Currie:2017eqf,Gehrmann-DeRidder:2019ibf,Czakon:2019tmo,Chen:2022tpk}.
The calculation of NNLO QCD corrections for $pp \rightarrow jjj$ production is among the most challenging and technically demanding computations undertaken to date~\cite{Czakon:2021mjy}. 
Reaching NNLO accuracy to higher final-state multiplicities remains at the frontier of current computational capabilities.
In $e^+e^-$ collisions, apart from dijet production, which is known up to next-to-next-to-next-to leading order (N$^3$LO) \cite{Chen:2025kez}, NNLO predictions are available only for 3-jet production \cite{Gehrmann-DeRidder:2007vsv,Weinzierl:2008iv,DelDuca:2016ily}, highlighting the need for further advancements.

In regions of the phase space characterised by a large hierarchy of scales, jet
observables become sensitive to the details of the QCD radiation. In such cases,
all-order resummation becomes
necessary to restore the predictive power of perturbation theory. While
general-purpose parton showers provide a way to resum such contributions to
all-order in perturbation theory, no general formalism exists for the
resummation of observables beyond next-to-leading-logarithmic (NLL) accuracy,
and only a relatively low number of observables are known at or beyond NNLL
accuracy~\cite{Jouttenus:2013hs,Arpino:2019ozn,Gao:2019ojf,Gao:2023ivm,Alioli:2023rxx,Chen:2023zlx,Gao:2024wcg,Cal:2024yjz} in processes with three or more partons at the Born level.

A major obstacle in extending the description of jet observables to higher
orders is the calculation of the NNLO ingredients entering the relevant
resummation formul\ae.
In many cases, observables obey factorisation theorems that allow the
organisation of the calculation by defining hard, soft, jet and beam functions.
The hard function encodes the dynamics at the hard scale $Q$ that characterises
the underlying process. The soft function accounts for the emission of soft
radiation from initial- and final-state partons. Beam functions describe the
evolution of collinear partons along the beam directions, while jet
functions capture the collinear dynamics of partons inside the jets.
This organisation of the perturbative series in terms of
  hard, soft and collinear functions remains useful even in the absence of a
factorisation theorem for the observable under
consideration. A notable example is the two-jet rate $y_{23}$ in
electron-positron collisions~\cite{Banfi:2016zlc}. Since deviations from a fully factorized ansatz can be systematically computed order by order in perturbation
theory, this strategy enables an efficient organisation of
the fixed-order expansion of the observable at higher perturbative orders.

In this work, we focus on the computation of jet
functions at next-to-next-to-leading order (NNLO).
In particular, we consider the class of observables that scale as a transverse
momentum in the limit where the radiation becomes soft and collinear to the jet
direction corresponding to the leg $\ell$, i.e.
\begin{equation}
  \label{eq:class}
\qf\sim k_t^{(\ell)}\, .
\end{equation} 
In the nomenclature of Soft-Collinear-Effective-Theory (SCET)~\cite{Bauer:2000yr,Bauer:2001yt,Bauer:2002nz,Beneke:2002ph,Beneke:2002ni}, observables scaling as~\eqref{eq:class} belong to the class of SCET$_{\rm II}$ variables.

In the case of observables like $N$-jettiness \cite{Stewart:2010tn} or the
transverse momentum of a colourless system (or heavy-quark pairs) and related
variables, the perturbative knowledge of beam
\cite{Catani:2011kr,Catani:2012qa,Gehrmann:2012ze,Ebert:2020unb,Ebert:2020yqt,Luo:2020epw,Abreu:2022zgo,Catani:2022sgr,Baranowski:2022vcn,Bell:2024epn},
soft
\cite{Angeles-Martinez:2018mqh,Abreu:2022sdc,Catani:2023tby,Bell:2023yso,Agarwal:2024gws,Baranowski:2024vxg,Liu:2024hfa}
and jet \cite{Becher:2006qw,Becher:2010pd,Bruser:2018rad,Banerjee:2018ozf}
functions is available at NNLO and beyond. On the contrary, for the
case of more general $k_T$-like variables in \eqref{eq:class}, which typically rely on the use of a jet clustering algorithm, the status of
  jet function calculations is less advanced due to the complications arising
  from the clustering history. A notable exception is inclusive jet production
  for the generalised $k_T$ jet algorithms. The two-loop anti-$k_T$ quark jet
  function for small jet radii has been
  computed in Ref.~\cite{Liu:2021xzi}. Aside from the TMD jet functions, which
  are partially known at NNLO~\cite{Gutierrez-Reyes:2019vbx,Fang:2024auf} and
  have also been employed for the variable defined in Ref.~\cite{Fu:2024fgj},
  state-of-the-art jet function calculations for generic $k_T$-like $n$-jet
  observables are currently limited to
  NLO~\cite{Banfi:2014sua,Buonocore:2022mle,Buonocore:2023rdw}.
It is worth emphasising that the knowledge of the beam, soft, and jet functions at a
given perturbative order is not only essential for all-order resummation, but
also enables fully differential fixed-order computations using slicing
methods~\cite{Catani:2007vq,Gaunt:2015pea}, thereby broadening the scope of
potential applications.

In this letter we present a semi-numerical strategy for
the computation of the NNLO quark jet function for a
$k_T$-like variable that smoothly captures the $n+1$ to $n$ jet transition in
$e^+e^-$ collisions. Our approach is fully general, and can be applied to any
observable in the class \eqref{eq:class}. We illustrate the main steps of our
method and we present explicit results for a variant of $y_{23}$ within the $E$-scheme and WTA \cite{Bertolini:2013iqa} scheme.
Related work has been independently presented in Refs.~\cite{Bell:2021dpb,Brune:2022cgr}.

It is well known that in the calculations of the perturbative jet, beam and soft
function for SCET$_{\rm II}$ observables one generally encounters divergent
integrals that are not regulated by dimensional regularisation
\cite{Collins:2008ht,Collins:2011zzd,Becher:2010tm,Echevarria:2011epo,Chiu:2012ir}.
Such divergences require the introduction of an additional {\it rapidity}
regulator, which is typically not necessary for SCET$_{\rm I}$ observables.
Several approaches to deal with this problem have been
formulated~\cite{Ji:2004wu,Chiu:2009yx,Chiu:2012ir,Becher:2011dz,Echevarria:2015byo};
in this paper we follow Ref.~\cite{Catani:2022sgr} and we regulate rapidity
divergences through the introduction of a {\it timelike} reference vector.

In combination with a general study of the factorisation properties for this
class of $k_T$-like observables \cite{inprep1}, the results we are going to
present constitute a building block for possible resummed computations for these
observables and are directly relevant for NNLO calculations of multijet cross
sections at high-energy colliders when these observables are used as slicing
variables \cite{inprep2}. Our results may also have a relevance in the matching
of fixed-order computations to parton shower simulations.

The paper is organised as follows. We introduce the framework for the
computation of jet functions for $k_T$-like observables in Sec.~\ref{sec:comp}.
In Sec.~\ref{sec:result} we give results for observables defined using the $k_T$
jet clustering algorithm for different jet recombination schemes. In
Sec.~\ref{sec:concl} we summarise our findings and we draw our conclusions.

\section{The computation}\label{sec:comp}

In our computation we regularise both ultraviolet (UV) and infrared (IR) divergences by using conventional
dimensional regularisation in $d=4-2\ep$ space-time dimensions. The
$\mathrm{SU}(N_c)$ QCD colour factors are $C_F=(N_c^2-1)/(2N_c)$, $C_A = N_c$, $T_R = 1/2$ and we consider $n_F$ massless quark flavours. 

The bare cumulative jet function for a resolution variable $\qf$ is defined as
\begin{align}
  \label{eq:defJ}
  \J^{ss^\prime}_{a}(\qfcut)=\theta(\qfcut)\delta^{ss^\prime} &+\sum_{n=2}^\infty \sum_{\Af_{n}}2(2\pi)^{d-1}\!\int\prod_{j=1}^n[dk_j]\delta\bigg(1-\sum_{j=1}^n z_j\bigg)\delta^{(d-2)}\!\bigg( \sum_{j=1}^n k_{j, \perp} \!\!\bigg)\nonumber \\
&\times  \frac{\calPhat^{ss^\prime}_{\Af_{n}}(k_1,k_2...k_n)}{S_{\Af_{n}}}
  \theta\!\Big( \qfcut -  \qft\Big) ,
\end{align}
where $a$ is the parton flavour ($a=q,{\bar q},g$), $\Af_{n}$ is a multi-index containing the flavours of the $n$ final-state QCD partons with momenta $k_1,k_2...k_n$, $\qft=\qft(k_1,k_2...k_n)$ is the collinear approximation of the resolution variable and $k_{j,\perp}$, $z_j$ are the (boost invariant) transverse momenta and momentum fractions involved in the multiparton collinear limit.
The factor $1/S_{\Af_n}$ in Eq.~\eqref{eq:defJ} is a symmetry factor for identical particles and
\begin{equation}
[dk_j]=\frac{d^dk_j}{(2\pi)^{d-1}}\delta_+(k_j^2)  
\end{equation}
is the phase space measure for massless particles.

The function $\calPhat_{\Af_n}(k_1,k_2...k_n)$ in Eq.~\eqref{eq:defJ} controls the limit of the QCD matrix element squared $|{\cal M}(k_1,k_2...k_n...)|^2$ in which $n$ final state partons become collinear
\begin{equation}
\left|{\cal M}_{a_1,a_2..a_n...}(k_1,k_2...k_n...)\right|^2\simeq {\cal T}_a^{ss^\prime}(p...) \calPhat_{a_1,a_2...a_n}^{ss^\prime} \, ,
\end{equation}
${\cal T}_a^{ss^\prime}(p...) $ being the spin polarisation tensor obtained by replacing the collinear partons with the parent parton $a$, which carries the flavour and momentum of the system $a_1+a_2+...a_n$ in the collinear limit. The function $\calPhat_{a_1,a_2...a_n}^{ss^\prime}$ is related to the customary collinear splitting kernels through the relation
\begin{equation}
 \calPhat_{a_1,a_2...a_n}^{ss^\prime} =\left(\frac{8\pi \as^u\mu_0^{2\ep}}{s_{1\dots n}}\right)^{n-1}{\hat P}^{ss'}_{a_1,a_2...a_n}\, ,
\end{equation}
where $s_{1\dots n}=(k_1+k_2+\dots k_n)^2$.
The functions ${\hat P}_{a_1,a_2...a_n}$ have the perturbative expansion in the bare coupling $\alpha_0$
\begin{equation}
{\hat P}^{ss'}_{a_1,a_2...a_n}=\sum_{l=0}^\infty \left(\frac{\alpha_0}{\pi}\right)^l\left(\frac{\mu^2}{s_{1...n}}\right)^{l \ep}{\hat P}_{a_1,a_2...a_n}^{(l)ss'} \, ,
\end{equation}
where ${\hat P}^{(0)}$ is the tree-level contribution, ${\hat P}^{(1)}$ the one-loop correction, and so forth.
The jet function ${\J}_a^{ss^\prime}(\qfcut)$ is in general a matrix in the spin indices $s$ and $s^\prime$ of the parton $a$. In the quark case, spin correlations are absent, and the matrix is diagonal
\begin{equation}
{\J}_q^{ss^\prime}(\qfcut)={\J}_q(\qfcut)\delta^{ss^\prime}\, .
\end{equation}
The perturbative expansion of ${\J}_{q}(\qfcut)$ reads
\begin{equation}
  \label{eq:pertJ}
 {\J}_{q}(\qfcut)=\theta(\qfcut) +\sum_{n=1}^\infty\left( \frac{\alpha_0}{\pi} \right)^n\left(\frac{\mu^2}{\qfcut^2}\right)^{n\ep }{\J}_{q}^{(n)}(\qfcut)\, .
\end{equation}
The bare coupling $\alpha_0=\alpha_0(\mu)$ is related to the (dimensionless) unrenormalised coupling $g_{\mathrm S}^u$ in the QCD Lagrangian by the relation
\begin{equation}
\as^u\mu_0^{2\ep} S_\ep=\alpha_0(\mu) \mu^{2\ep}  
\end{equation}
with $\asbare=(g_{\mathrm S}^u)^2/(4\pi)$,
and $S_\ep$ is the customary spherical factor $S_\ep=(4\pi)^\ep e^{-\ep \gamma_E}$.

The computation of the jet function in Eq.~\eqref{eq:defJ} is generally affected by rapidity divergences \cite{Collins:2008ht,Collins:2011zzd,Becher:2010tm,Echevarria:2011epo,Chiu:2012ir}, which manifest themselves as singularities as $z_i\to 0$ that are not regulated by dimensional regularisation.
In this paper we treat rapidity divergences by using a {\it time-like} auxiliary vector $N$ to define the momentum fractions. This regularisation scheme, that we dub ``$z_N$ prescription'', has been introduced in Ref.~\cite{Catani:2022sgr}\footnote{A similar (though not exactly equivalent) regularisation procedure was used in Ref.~\cite{Ji:2004wu}, whereas Ref.~\cite{Collins:2011zzd} uses a {\it space-like} auxiliary vector $N$.}.
The momentum fraction $z_N$ for a parton with momentum $k$ reads
\begin{equation}
  \label{eq:zN}
z_N=\frac{k\cdot N}{p\cdot N}=z+\frac{N^2 k_\perp^2}{(2p\cdot N)^2 z}   \, ,
\end{equation}
where $p$ is the collinear direction and $k_\perp$ is the transverse momentum
involved in the collinear splitting. In the strict collinear limit the term
proportional to $k_\perp^2$ in Eq.~\eqref{eq:zN} is subleading. The presence of $N^2>0$ in Eq.~\eqref{eq:zN} prevents factors $1/z_N$ in the splitting kernels from becoming singular and ensures that the integrals in Eq.~\eqref{eq:defJ} are well-defined.
Due to the $z_{N}$ prescription, the integral associated with the soft endpoint
  of the splitting kernel is no longer scaleless. As a result, the procedure
  generates a non-vanishing {\it zero-bin} contribution, which we combine with the
  soft function~\cite{inprep1} to avoid double counting. We note that in our calculation the replacement $z\to z_N$ is
implemented only in the singular term of the relevant splitting kernels, as
described below. We denote the perturbative coefficients of the bare jet function with this regularisation by $\J^{(n)}_{N,q}$.

We start from the NLO computation. The only quark-initiated splitting at this order is $q\to gq$. The relevant splitting kernel is
\begin{equation}
{\hat P}^{(0)}_{gq}(z)=C_F\left[\frac{1+(1-z)^2}{z}-\ep z\right]\, .
\end{equation}
The regularised kernel ${\hat P}^{(0)}_{N,gq}(z)$ is obtained from ${\hat P}^{(0)}_{gq}(z)$ by isolating the $1/z$ singular term and replacing $z$ with $z_N$ therein:
\begin{equation}
\label{eq:PNgq}
{\hat P}^{(0)}_{N,gq}(z)=C_F\left[\frac{2}{z_N}-2+(1-\ep)z\right]\, .
\end{equation}
From Eqs.~\eqref{eq:defJ} and \eqref{eq:pertJ} we have
\begin{equation}
  \label{eq:J1int}
  \J^{(1)}_{N,q}(\qfcut ) = 4\, \qfcut^{2\eps}\frac{e^{\eps \gamma _E}}{\Gamma(1-\eps)}\! \int \!
  d\Phi_2^{(c)}
  \frac{{\hat P}^{(0)}_{N,gq}(z_1)}{2 k_1\cdot k_2} \theta (\qfcut - \qft)\, ,
\end{equation}
where we defined the collinear phase space
\begin{equation}
 d\Phi_n^{(c)} =\frac{1}{\left( \Omega_{2-2\eps} \right)^{n-1}}\bigg(\prod_{j=1}^n d^dk_j \delta _+(k_j^2)\bigg)  \delta(1-\sum_{j=1}^n z_j) \delta^{(d-2)} (\sum_{j=1}^n\vec{k}_{j,\perp}) \, .
\end{equation}
In the following, we focus on the particular case where the variable fulfils the condition $\qft=k_\perp$~\footnote{This assumption is not essential for the construction of the method but leads to some useful simplifications.} in the two-parton collinear limit. Then, we can further manipulate
the previous equation as
\begin{equation}
  \J^{(1)}_{N,q}(\qfcut ) =\qfcut^{2\eps}\frac{e^{\eps \gamma _E}}{\Gamma(1-\eps)}\int_0^1 dz \int_0^{\qfcut} dk_\perp k_\perp^{-1-2\ep} {\hat P}^{(0)}_{N,gq}(z)\, .
\end{equation}
The final integration can be carried out by using the identity
\begin{equation}
\label{eq:identity}  
\frac{1}{z_N}=\left(\frac{1}{z}\right)_+-\frac{1}{2}\delta(z)\ln\frac{N^2 k^2_\perp}{(2p\cdot N)^2}+{\cal O}\left(\frac{k_\perp \sqrt{N^2}}{2p\cdot N}\right)  
\end{equation}  
and we obtain
\begin{equation}\label{eq:jetfunNLO}
{\J}_{N, q}^{(1)}=\frac{e^{\ep\gamma_E}}{\Gamma(1-\ep)} C_F\left[\frac{1}{2\ep^2}+\frac{L_N}{\ep}+\frac{3}{4\ep}+\frac{1}{4}\right]\, ,
\end{equation}
where we have defined
\begin{equation}
  \label{eq:LN}
  L_N=\ln\left(\frac{\sqrt{N^2}\qfcut}{2p\cdot N}\right)\, .
\end{equation}
At NNLO the quark jet function receives contributions from the $q\to gq$ splitting at one-loop order and from all the possible tree-level triple collinear splittings $q\to {\bar q}^\prime q^\prime q$, $q\to {\bar q}qq$, $q\to ggq$.  The one-loop contribution is controlled by the $q\to gq$ splitting kernel, which reads \cite{Bern:1999ry,Kosower:1999rx}
\begin{align}
\label{eq:P1}
{\hat P}_{gq}^{(1)}(z)&=\frac{1}{2}c_P\cos(\pi\epsilon)\Bigg\{C_F(C_A-C_F)\frac{1-\epsilon z}{1-2\epsilon}
  +{\hat P}^{(0)}_{gq}(z)\left(-\frac{1}{\epsilon^2}\right)\Bigg[C_A\,{}_2F_1\left(1,-\ep,1-\ep,1-\frac{1}{z}\right)\nonumber\\
  &-2(C_F-\frac{1}{2}C_A)\left(1-{}_2F_1\left(1,-\ep,1-\ep,-\frac{z}{1-z}\right)
  \right)\Bigg]\Bigg\}\, ,
\end{align}
where
\begin{equation}
c_P=e^{\ep\gamma_E}\frac{\Gamma^2(1-\ep)\Gamma(1+\ep)}{\Gamma(1-2\ep)}=1-\frac{\pi^2}{12}\ep^2-\frac{7}{3}\zeta_3 \ep^3-\frac{47}{1440}\pi^4 \ep^4+{\cal O}(\ep^5)\, .
\end{equation}
The regularised version of the one-loop splitting kernel ${\hat P}^{(1)}_{N,gq}(z)$ is obtained by replacing the tree level kernel ${\hat P}^{(0)}_{gq}(z)$ in the first line of Eq.~\eqref{eq:P1} with ${\hat P}^{(0)}_{N,gq}(z)$.
The ensuing contribution to the jet function can be straightforwardly evaluated by replacing the tree-level kernel ${\hat P}^{(0)}_{N,gq}(z)$ with its one-loop correction ${\hat P}^{(1)}_{N,gq}(z)$ in Eq.~\eqref{eq:J1int}.

The computation of the tree-level contributions to $\J^{(2)}_{N,q}$ is driven by the $1\to 3$ collinear splittings \cite{Campbell:1997hg,Catani:1998nv,Catani:1999ss}.
The collinear approximation is now defined in terms of two momentum
  fractions $z_{i}$ and $z_{j}$. Consequently, the structure of rapidity
  divergences becomes more intricate, and the extension of the $z_{N}$
  prescription correspondingly more subtle. Our strategy relies on analysing the
  divergences that can occur in the splitting kernels, i.e., at the level of the
  squared matrix elements. For configurations characterised by strongly-ordered
  independent emissions, rapidity divergences arise in the limit where one or
  both momentum fractions vanish, $z_{i}\to 0$ and/or $z_{j}\to 0$. In contrast,
  for configurations involving correlated emissions, rapidity divergences occur
  only when the sum of the momentum fractions vanishes, $z_{i}+z_{j}\to 0$. The
  former case displays a more intricate structure due to the overlap of the two
  limits. Since we only require that the $z_{N}$ replacement yields finite
  integrals, a certain freedom remains in the precise definition of the regularised kernels, which is compensated by
  different zero-bin contributions.
  In the following, we present our analysis for all the $1\to 3$ splittings relevant to the
  quark jet function, and provide the explicit expressions of the corresponding $z_{N}$-regularised kernels.

We start from the different flavour $q\to {\bar q}'q'q$ term. The corresponding splitting kernel reads
\begin{equation}
    \label{eq:Pqqpqpx}
    {\hat P}^{(0)}_{\bar{q}'_1  q'_2 q_3} =  \frac{1}{2}C_FT_R \frac{s_{123}}{s_{12}} \bigg[\bigg(  -\frac{t_{12,3}^2}{s_{12}s_{123}} + \frac{4 z_3 + (z_1-z_2)^2}{z_1+z_2}\bigg)+ (1-2\eps) \bigg( z_1+z_2 - \frac{s_{12}}{s_{123}}\bigg) \bigg]\, ,
\end{equation}
where
\begin{equation}
t_{ij,k}=2\frac{z_i s_{jk}-z_j s_{ik}}{z_i+z_j}+\frac{z_i-z_j}{z_i+z_j}s_{ij}\, .
\end{equation}
In this case the rapidity divergence appears only when the sum of the quark and antiquark momentum fractions $z_1+z_2$ vanishes. The modified splitting kernel is obtained from Eq.~\eqref{eq:Pqqpqpx} as
\begin{align}
    \label{eq:qqpqpx}
    {\hat P}^{(0)}_{N,\bar{q}'_1  q'_2 q_3} = & \frac{1}{2}C_FT_R \frac{s_{123}}{s_{12}} \bigg[ \frac{1}{z_{1,N} + z_{2,N}}\bigg(  -\frac{(z_1+z_2)t_{12,3}^2}{s_{12}s_{123}} + 4 z_3 + (z_1-z_2)^2\bigg) \nonumber \\ 
    & + (1-2\eps) \bigg( z_1+z_2 - \frac{s_{12}}{s_{123}}\bigg) \bigg].
\end{align}
The related contribution to the jet function is
\begin{equation}\label{eq:jetCFTRnf}
  {\J}_{N, q}^{(2)}(\qfcut)\Big|_{C_FT_R}=8 \qfcut^{4\epsilon}\frac{e^{2\ep\gamma_E}}{\Gamma^2(1-\ep)}\int d\Phi_3^{(c)}\frac{{\hat P}^{(0)}_{N,\bar{q}'_1q'_2q_3}}{s_{123}^2} \theta (\qfcut-\qft)\, .
\end{equation}
We find it useful to employ the variables defined in Ref.~\cite{Bell:2022nrj}:
\begin{equation}
    \label{eq: Bell variables}
    a = \frac{z_1 k_{2,\perp}}{z_2k_{1,\perp}}, \quad b=\frac{k_{1,\perp}}{k_{2,\perp}}, \quad z=z_1+z_2,\quad m_T = \sqrt{z\left(\frac{k_{1,\perp}^2}{z_1} + \frac{k_{2,\perp}^2}{z_2} \right)}, \quad x_{12} = \frac{1-\cos \varphi _{12}}{2} \, . 
\end{equation}
The variable $a$ represents the rapidity difference between $k_1$ and $k_2$,
$z$ is the momentum fraction of the momentum $k_1+k_2$, while $m_T$ is its transverse mass, and $\varphi _{12}$ is the angle between the $(d-2)$-dimensional vectors $\vec{k}_{1,\perp}$ and $\vec{k}_{2,\perp}$.
The parametrization has only one dimensionful variable, $m_T$, and $\qft$ can be written as
\begin{equation}
    \label{eq:qft-coll}
    \qft^2 = m^2_T F(a,b,z,x_{12})\, ,
\end{equation}
where the dimensionless function $F(a,b,z,x_{12})$ specifies the observable. In the new variables, we can write 
\begin{equation}
 z_{1,N} + z_{2,N}=z+\frac{m_T^2 N^2}{z(2p\cdot N)^2}\, ,
\end{equation} and we can use Eq.~\eqref{eq:identity} to extract the leading-power contribution from the regularised splitting kernel.
The integration over the dimensionful variable $m_T$ is straightforward, and the rest of the calculation can be organised by remapping the remaining variables onto the unit hypercube, with suitable changes of variables to avoid overlapping singularities. At this point, the integrand can be expanded in $\ep$ and the final integrations are carried out numerically.

The contribution of the $q\to {\bar q}qq$ splitting is obtained from the $q\to {\bar q}'q'q$ by adding the interference term
\begin{equation}
    {\hat P}^{(0)}_{\bar{q}_1 q_2 q_3} = [{\hat P}^{(0)}_{\bar{q}'_1  q'_2 q_3 } + (2\leftrightarrow 3)] + {\hat P}_{\bar{q}_1  q_2 q_3 }^{(0)(\text{id})} \, ,
\end{equation}
where
\begin{align}
\label{eq:qqq}
    {\hat P}_{\bar{q}_1q_2q_3}^{(0)(\text{id})} & = C_F\left(C_F -\frac{1}{2}C_A \right)\left\{ (1-\eps) \left(\frac{2 s_{23}}{s_{12}}-\eps\right)  \right. \\
    & +\frac{s_{123}}{s_{12}}\left[ \frac{1+z_1^2}{1-z_2}-\frac{2z_2}{1-z_3} - \eps \left(\frac{(1-z_3)^2}{1-z_2} +1+z_1-\frac{2z_2}{1-z_3}\right)-\eps^2 (1-z_3) \right] \\
    & \left. -\frac{s_{123}^2}{s_{12}s_{13}}\frac{z_1}{2}\left[\frac{1+z_1^2}{(1-z_2)(1-z_3)} - \eps \left( 1+2\frac{1-z_2}{1-z_3} \right) - \eps ^2 \right] \right\} + (2\leftrightarrow 3)\, .
\end{align}
The splitting kernel $ {\hat P}_{\bar{q}_1q_2q_3}^{(0)(\text{id})}$ does not exhibit rapidity divergences and therefore we do not define a $z_N$-regularised version of it.
The computation can be carried out with the parametrisation of Eq.~\eqref{eq: Bell variables} with no additional complications.

We now move to the $q\to ggq$ contribution, which can be organised into an abelian and a non-abelian part:
\begin{equation}
    {\hat P}^{(0)}_{g_1 g_2 q_3} = C_F^2 {\hat P}_{g_1 g_2 q_3}^{(0)(\text{ab})} + C_F C_A {\hat P}_{g_1 g_2 q_3}^{(0)(\text{nab})}\, .
\end{equation}
The non-abelian $q\to ggq$ kernel reads
\begin{equation}
\begin{split}
\label{eq: q to qqg non-abelian}
        P^{(0)(\rm nab)}_{g_1g_2q_3} &= \left\{(1-\eps)\left( \frac{t_{12,3}^2}{4s_{12}^2} + \frac{1}{4} - \frac{\eps}{2}\right) + \frac{s_{123}^2}{2s_{12}s_{13}} \left[\frac{(1-z_3)^2(1-\eps) + 2z_3}{z_2} \right. \right. \\
        & \left. +\frac{z_2^2(1-\eps)+2(1-z_2)}{1-z_3} \right] - \frac{s_{123}^2}{4s_{13}s_{23}}z_3 \left[ \frac{(1-z_3)^2(1-\eps)+2z_3}{z_1z_2} + \eps(1-\eps)  \right]\\
        & + \frac{s_{123}}{2s_{12}} \left[ (1-\eps) \frac{z_1(2-2z_1+z_1^2)-z_2(6-6z_2+z_2^2)}{z_2(1-z_3)} + 2\eps \frac{z_3(z_1-2z_2)-z_2}{z_2(1-z_3)}  \right] \\
        & + \frac{s_{123}}{2s_{13}} \left[ (1-\eps) \frac{(1-z_2)^3+z_3^2-z_2}{z_2(1-z_3)} - \eps \left( \frac{2(1-z_2)(z_2-z_3)}{z_2(1-z_3)}-z_1+z_2 \right) \right. \\
        & \left. \left. -\frac{z_3(1-z_1)+(1-z_2)^3}{z_1z_2} + \eps (1-z_2)\left( \frac{z_1^2+z_2^2}{z_1z_2}-\eps \right) \right] \right\} + (1\leftrightarrow 2)\, .
\end{split}
\end{equation}
In this case, like the $q\to \bar{q}'q'q$ case, the emissions are correlated, and the rapidity divergence appears only when the sum $z_1+z_2$ vanishes. Therefore, we define the $z_N$-regularised version of this splitting kernel by only modifying the terms that are singular when the sum of the two momentum fraction vanishes: 
\begin{equation}
\begin{split}
    \label{eq: q to qqg non-abelian zn}
    & {\hat P}_{N,g_1g_2q_3}^{(0)\rm{(nab)}}={\hat P}_{g_1g_2q_3}^{(0)\rm{(nab)}} - \bigg(1-\frac{z_1+z_2}{z_{N,1}+z_{N,2}} \bigg) \bigg[ \frac{(1-\eps)(z_1s_{23}-z_2s_{13})^2}{(z_1+z_2)^2 s_{12}^2}\\
    & +\frac{(z_1+2z_2)s_{123}^2}{z_2(z_1+z_2)s_{12}s_{13}}-\frac{s_{123}^2}{2z_1 z_2 s_{13} s_{23}} + \frac{(z_1-3z_2)s_{123}}{z_2(z_1+z_2)s_{12}} - \frac{s_{123}}{z_1(z_1+z_2)s_{13}} + (1 \leftrightarrow 2)\bigg]\, .
\end{split}
\end{equation} 
The corresponding calculation can be carried out similarly to what was done for $C_F T_R$.
The only difference is that in this case there is an additional pole due to the configuration in which one of the two emitted gluons is soft. This configuration, however, does not lead to additional complications. 

The most intricate part of the computation is the evaluation of the abelian $q\to ggq$ contribution. 
The corresponding splitting kernel is: 
\begin{equation}
\begin{split}
    {\hat P}_{g_1 g_2 q_3}^{(0)(\text{ab})} =&   \left\{\frac{s_{123}^2}{2s_{13}s_{23}}z_3 \left[\frac{1+z_3^2}{z_1z_2} -\eps \frac{z_1^2+z_2^2}{z_1z_2}-
    \eps(1+\eps)\right]\right.\\
    & +\frac{s_{123}}{s_{13}}\left[\frac{z_3(1-z_1)+(1-z_2)^3}{z_1z_2}+\eps ^2 (1+z_3)-\eps (z_1^2+z_1z_2+z_2^2)\frac{1-z_2}{z_1z_2} \right]\\
    &\left. +(1-\eps) \left[\eps -(1-\eps) \frac{s_{23}}{s_{13}}\right]\right\} + (1\leftrightarrow 2)\, .
\end{split}
\end{equation}
In this case the emitted gluons are uncorrelated, and the rapidity divergences appear also when only one of the two momentum fractions $z_1$ and $z_2$ vanishes. To define the $z_N$-regularised version of the abelian $q\to ggq$ splitting, we isolate its strongly-ordered limit by defining: 
\begin{equation}
\label{eq: decomposition of CFCF kernel}
    {\hat P}_{g_1 g_2 q_3}^{(0)(\rm{ab})} = {\hat P}_{g_1 g_2 q_3}^{(0)\rm{(ab),S.O.}}+{\hat P}_{g_2 g_1 q_3}^{(0)\rm{(ab),S.O.}} + R_{g_1 g_2 q_3}^{(0)(\rm{ab})} \, , 
\end{equation}
where the strongly-ordered term reads
\begin{equation}
    {\hat P}_{g_1 g_2 q_3}^{(0)\rm{(ab),S.O.}} =\frac{s_{123}}{s_{23}} {\hat P}^{(0)}_{gq}(z_1){\hat P}^{(0)}_{gq}\bigg(\frac{z_2}{z_2+z_3}\bigg) \, .
\end{equation}
The remainder $R_{g_1 g_2 q_3}^{(0)(\rm{ab})}$ does not feature rapidity divergences and therefore does not need regularisation. The strongly-ordered contribution ${\hat P}_{g_1 g_2 q_3}^{(0)\rm{(ab),S.O.}}$ is regularised by substituting the NLO splitting kernels with their $z_N$ versions:
\begin{equation}
    {\hat P}_{N,g_1 g_2 q_3}^{(0)\rm{(ab),S.O.}} =\frac{s_{123}}{s_{23}} {\hat P}^{(0)}_{N,gq}(z_1){\hat P}^{(0)}_{N,gq}\bigg(\frac{z_2}{z_2+z_3}\bigg)  \, . 
\end{equation}
The contribution coming from the remainder $R_{g_1g_2q_3}^{(0)(\rm{ab})}$ can be computed with the strategy employed for the $C_F T_R$ 
and $C_F C_A$ terms. The most complicated part of our calculation is the integration of the strongly-ordered term. For this contribution we find it useful to employ the following phase space parametrization. We introduce the variables
\begin{equation}
  \tilde{z}_2 = \frac{z_2}{z_2+z_3}, \quad \vec{k}_{23,\perp} = \frac{z_3 \vec{k}_{2,\perp} - z_2 \vec{k}_{3,\perp}}{z_2+z_3} \, . 
\end{equation} 
The variable $\tilde{z}_2$ is the momentum fraction associated with the splitting $q\to g_2 q_3$ and $\vec{k}_{23,\perp}$ is the relative transverse momentum of the partons 2 and 3, divided by their total momentum fraction. 
With these variables, we can write 
\begin{equation}
\label{eq:z1z2}  
  z_{N,1}=z_1+\frac{N^2 k_{1,\perp}^2}{(2p\cdot N)^2 z_1}\quad \tilde{z}_{N,2}=\tilde{z}_{2}+\frac{N^2 k_{23,\perp}^2}{(2p\cdot N)^2 \left( 1-z_1 \right)^2\tilde{z}_2}+\dots\, ,
\end{equation}
where the dots represent terms that are subleading in all relevant limits.
We write the phase space in terms of the variables
\begin{align}
  & z_1, \quad \tilde{z}_2,\quad y \equiv \min\bigg\{\frac{k_{1,\perp}}{k_{23,\perp}}, \frac{k_{23,\perp}}{k_{1,\perp}}\bigg\}, \quad k_\perp \equiv \max \{k_{1,\perp},k_{23,\perp}\}, \quad\cos \varphi \equiv \frac{\vec{k}_{1,\perp}\cdot \vec{k}_{23,\perp}}{k_{1,\perp} k_{23,\perp}}\, .
\end{align}
The main difficulty in integrating the strongly-ordered splitting function arises from the presence of rapidity divergences in the individual limits $z_1 \to 0$ and $z_2 \to 0$, not only in the combined limit $z_1 + z_2 \to 0$. To overcome this complication, we divide the integration of $P_{N,g_1g_2q_3}^{(0),\rm {(ab)S.O.}}$ into two pieces: 
\begin{equation}
  \label{eq:subpiece}
  \int d\Phi _3^{(c)}\frac{P_{N,\,g_1g_2q_3}^{(0),\rm (ab)S.O.}}{s_{123}^2}\theta (\qfcut - \qft) = \int d\Phi _3^{(c)}\frac{P_{N,\,g_1g_2q_3}^{(0),\rm (ab)S.O.}}{s_{123}^2}\theta (\qfcut - k_\perp) + \int d\Phi _3^{(c)}\frac{P_{N,\,g_1g_2q_3}^{(0),\rm (ab)S.O.}}{s_{123}^2}\theta _{\rm sub}\, ,
\end{equation}
where we defined the subtracted Heaviside function 
\begin{equation}
  \theta _{\rm sub} = \theta (\qfcut - \qft) - \theta (\qfcut - k_\perp) \, .
\end{equation}
We refer to the first term on the right-hand side of Eq.~\eqref{eq:subpiece} as the {\it endpoint term}, and to the second as the {\it subtracted term}. The endpoint term is common to all variables belonging to the class considered in this paper. The subtracted term is free of poles, since $\theta _{\rm sub}$ acts as a counterterm that regulates all the singularities of $P_{N,\,g_1g_2q_3}^{(0),\rm {(ab)S.O.}}$, but still contains terms proportional to powers of $L_N$.

To integrate the subtracted term, we separate contributions that contain at most one regularised momentum fraction, and contributions that contain the product $1/(z_{1,N}\tilde{z}_{2,N})$.
In the integrals in which only one of the momentum fractions is replaced with its $z_N$ version, the calculation can be carried out by using identities similar to Eq.~\eqref{eq:identity}.
In the integrals in which both momentum fractions need to be regularised by the $z_N$ prescription, we cannot simply use Eq.~\eqref{eq:identity} on both momentum fractions separately. Instead, we need to perform a uniform expansion of the product $1/(z_{1,N}\tilde{z}_{2,N})$ in the limit in which both gluons become soft. We provide the relevant expansion in Appendix~\ref{app:distributional} in Eq.~\eqref{eq:z1Nz2NDistribution}.

The endpoint term is the only one leading to $\ep$ poles. Furthermore, it contains additional divergences in the strongly-ordered limit, $y \to 0$, that overlap with the rapidity divergences. This prevents us from directly applying the expansions in Eqs.~\eqref{eq:identity} and \eqref{eq:z1Nz2NDistribution}. However, since the $\theta$-function in the endpoint term cuts on $k_\perp$ rather than $\qft$, we can calculate the endpoint term once and for all using for example sector decomposition techniques \cite{Binoth:2003ak}. The explicit result for the endpoint contribution is presented in Sect.~\ref{sec:result}.

\section{Results}\label{sec:result}

We apply the general strategy outlined in the previous section to the
  computation of the jet function for a generic $n$-jet resolution variable
  defined by using a recursive recombination jet algorithm with the distance
\begin{equation}
    d_{ij}^2 = \frac{E_i^2 E_j^2}{(E_i+E_j)^2} 2(1-\cos \theta _{ij})\, .
\end{equation}
This is a possible variant of the distance used in the $k_T$ algorithm for
$e^+e^-$ collisions \cite{Catani:1991hj}, which makes the calculations slightly
simpler. In particular, for a single emission the variable approaches the
transverse momentum relative to the jet direction in the two-parton collinear
limit, and, thus, at NLO the jet function is given by
Eq.~\eqref{eq:jetfunNLO}~\footnote{The $z_{N}$-regulated NLO jet function for
  the case of the standard distance used in the $k_T$ algorithm can be found
  in Ref.~\cite{Buonocore:2023rdw}.}. We run the recursive
recombination on a generic $n+k$ parton system until $n+1$ pseudopartons are
left. We then define $\qf=\min\{d_{ij}\}$. We consider two recombination
schemes:
\begin{itemize}
    \item $E$-scheme, where the momentum $k_{ij}$ obtained recombining particles with momenta $k_i$ and $k_j$ is
    \begin{equation}
        k_{ij} = k_i + k_j\, .
    \end{equation}
    \item Winner-take-all (WTA) scheme \cite{Bertolini:2013iqa}, where  the momentum of the recombined particle is
    \begin{equation}
        k_{ij} = (E_i+E_j)\left(\frac{k_i}{E_i}\theta(E_i-E_j) + \frac{k_j}{E_j} \theta (E_j-E_i)\right)\, .
    \end{equation}
\end{itemize}
In the $E$-scheme, the recombined momentum is simply the sum of the original
momenta, and in general it is massive. In the WTA scheme, the recombined momentum
is massless and it has the direction of the more energetic pseudo-parton.
The variable $\qf$ belongs to the general class of transverse-momentum variables
considered in Ref.~\cite{Buonocore:2022mle}, and dubbed $\ktness$.

In the parametrization of Eq.~\eqref{eq: Bell variables}, the dimensionless observable function $F(a,b,z,x_{12})$ reads
\begin{equation}
  F(a,b,z,x_{12}) = \Theta _{12} F_{12}(a,b,z,x_{12})+\Theta _{13} F_{13}(a,b,z,x_{12})+\Theta _{23} F_{23}(a,b,z,x_{23})\, ,
\end{equation}
where $\Theta _{ij}\equiv\theta (d_{ik}-d_{ij})\theta (d_{jk}-d_{ij})$ specifies the phase space region in
which partons $i$ and $j$ are clustered first.
In the $E$-scheme, the functions $F_{ij}(a,b,z,x_{12})$ read
\begin{align}
  & F_{12}(a,b,z,x_{12}) = \frac{a[(1+b)^2-4bx_{12}]}{(a+b)(1+ab)}\, , \quad F_{13}(a,b,z,x_{12}) = \frac{a}{(a+b)(1+ab)}\, , \nonumber \\
  & F_{23}(a,b,z,x_{12}) = \frac{ab^2}{(a+b)(1+ab)} \, .
\end{align}
In the WTA scheme, they are given by
\begin{align}
  F_{12}^{\rm WTA}(a,b,z,x_{12}) &=  \frac{a}{(a+b)(1+ab)}\Big[ (1+ab)^2 - 2b(1+ab)z(-1+a+2x_{12})\nonumber\\
    & ~~~~~~~~~~~~~~~~~~~~~~~+ b^2 \big((1-a)^2 + 4 a x_{12}\big) z^2    \Big]\, , \nonumber \\
    F_{13}^{\rm WTA}(a,b,z,x_{12}) &=  \frac{(1+ab-z)^2}{a(a+b)(1+ab)^5(1-z)^2} \Bigg\{ \bigg[ a^2 (1+a^2 b^2(1-z)^2  + 2 b z -4 b x_{12} z  \nonumber \\
    &  + b^2 z^2 + 2 a b (z-1) (-1+b(-1+2x_{12})z) \bigg] \theta \big( 1+ab-z-2abz \big)  \nonumber \\
    & + \bigg[(1+ab)^2 \big(1+a^2 + a(-2+4x_{12})\big) (1-z)^2 \bigg] \theta \big(-1-ab+z+2abz\big) \Bigg\}\, , \nonumber \\
    F_{23}^{\rm WTA}(a,b,z,x_{12}) = & \frac{ab^2(-1+ab(-1+z))^2}{(a+b)(1+ab)^5(1-z)^2} \Bigg\{ \bigg[(1+ab)^2 (1+a^2 +a(-2+4x_{12}))(1-z)^2 \bigg] \times \nonumber \\
    & \theta \big(-1-ab+2z+abz\big) + \bigg[ (-1+z)^2 -2a(-1+z)(b+z-2x_{12}z) + \nonumber \\
    & + a^2 (b^2 + b(2-4x_{12})z + z^2 ) \bigg] \theta \big(1+ab-2z-abz \big) \Bigg\}\, .
\end{align}
We are now ready to present our results. The NNLO quark jet function can be written as
\begin{equation}
 {\J}_{N,q}^{(2)} = L_N^2 \sum_{k=0}^2 \frac{D_k}{\eps^k} + L_N \sum _{k=0}^3 \frac{A_k}{\eps ^k} + \sum _{k=0}^4 \frac{B_k}{\eps ^k}\, , 
\end{equation}
where $L_N$ is defined in Eq.~\eqref{eq:LN}.
The coefficients $D_k$, $A_k$ and $B_k$ for $k\geq 1$ do not depend on the recombination scheme and read
\begin{align}
    & D_2 = \frac{C_F^2}{2}, \quad D_1 = 0, \quad A_3 = \frac{C_F^2}{2},  \quad A_2 = \frac{3}{4}C_F^2 + \frac{11}{24}C_F C_A -\frac{1}{6}C_F n_F T_R, \nonumber \\
    & A_1 = \bigg(\frac{1}{4}-\frac{\pi^2}{12}\bigg) C_F^2 + \bigg( \frac{67}{72}-\frac{\pi^2}{24} \bigg)C_F C_A -\frac{5}{18} C_F n_F T_R,  \nonumber \\
    & B_4 = \frac{C_F^2}{8}, \quad B_3 = \frac{3}{8}C_F^2 + \frac{11}{96}C_F C_A -\frac{1}{24} C_F n_F T_R,  \nonumber \\
    & B_2 = \bigg(\frac{13}{32}-\frac{\pi^2}{48}\bigg) C_F^2 + \bigg(\frac{83}{144}-\frac{\pi^2}{96}\bigg)C_F C_A -\frac{7}{36} C_F n_F T_R, \nonumber \\
    & B_1 = \bigg(\frac{15}{64}-\frac{\pi^2}{8} + \frac{2}{3}\zeta _3\bigg) C_F^2 + \bigg(\frac{1357}{1728} + \frac{11}{576}\pi^2 -\frac{13}{16}\zeta_3\bigg)C_F C_A + \bigg(-\frac{101}{432} -\frac{\pi^2}{144}\bigg)C_F n_F T_R\, .
\end{align}
The coefficients $A_0$, $B_0$ and $D_0$ were computed numerically. In the $E$-scheme they read
\begin{align}
  \label{eq:res1}
 & D_0^E = -1.40471(2)\, C_F^2, \nonumber \\
    & A_0^E =  -0.17976(1)~C_F^2 -2.20169(6)~C_F C_A -0.12794(2)~C_F n_F T_R\, ,\nonumber \\
    & B_0^E = 4.514(1)~C_F^2 -0.3084(6)~C_F C_A -0.2210(1)~C_F n_F T_R\, ,
\end{align}
and in the WTA scheme we find
\begin{align}\label{eq:res2}
    & D_0^{\rm WTA} = -0.8224671(5)\, C_F^2, \nonumber \\
    & A_0^{\rm WTA} = 2.01322(9)~C_F^2 -2.64831(2)~C_F C_A -0.0766(1)~C_F n_F T_R, \nonumber \\
    & B_0^{\rm WTA} = 8.3346(8)~C_F^2 -1.7774(7)~C_F C_A -0.0733(2)~C_F n_F T_R \, .
\end{align}
For completeness, we finally report also the corresponding coefficients for the observable-independent endpoint contribution,
i.e., the contribution to the jet function from the first term in Eq.~(\ref{eq:subpiece}).
They read
\begin{equation}
  \begin{aligned}\label{eq:res3}
   &D_{2}^{\mathrm{EP}}=\frac{1}{2} C_F^2\, ,\quad D_{1}^{\mathrm{EP}}=0\, ,\quad D_{0}^{\mathrm{EP}}=-0.8224671(5)~C_F^2\, ,\quad A_{3}^{\mathrm{EP}}=\frac{1}{2}C_F^2\, ,\quad A_{2}^{\mathrm{EP}}=\frac{3}{4}C_F^2\, ,\\
        &A_{1}^{\mathrm{EP}}=C_F^2\left(\frac{1}{4}-\frac{\pi ^2}{12}\right)\, ,\quad A_{0}^{\mathrm{EP}}=-1.6343868(8)~C_F^2\, ,\quad B_{4}^{\mathrm{EP}}=\frac{1}{8}C_F^2\, ,\quad B_{3}^{\mathrm{EP}}=\frac{3}{8}C_F^2\, ,\\
        &B_{2}^{\mathrm{EP}}=C_F^2\left(\frac{23}{32}-\frac{5 \pi ^2}{48}\right)\, ,\quad B_{1}^{\mathrm{EP}}=0.06315(3)~C_F^2\, ,\quad B_{0}^{\mathrm{EP}}= 0.4688(1)~C_F^2 \, .
  \end{aligned}
\end{equation}

In Eqs.~(\ref{eq:res1}-\ref{eq:res3}), alongside the numerical results, we quote the integration error.
The observable-dependent subtracted contributions, which involve four-dimensional integrals, are computed with a dedicated Fortran code using {\sc Cuba} \cite{Hahn:2004fe}, while the endpoint contribution coefficients are evaluated with \texttt{Mathematica}.


\section{Summary}\label{sec:concl}

The precise description of jet production processes requires observables capable of efficiently
encoding the dynamics of the energy flow in hadronic final states.
Transverse-momentum ($k_T$) like observables are shape variables that scale as the transverse momentum in the limit where the radiation becomes soft and collinear to the jet direction, thereby naturally capturing the leading singular behaviour of the multijet matrix elements.

In the region where additional radiation is inhibited, the multijet cross section is described in terms of hard, soft, beam, and jet functions. This organisation of the perturbative series remains useful even when a factorisation theorem for the observable under consideration is not available. Jet functions describe the collinear dynamics of partons inside a jet, and their
perturbative evaluation for $k_T$-like variables is complicated by the necessity to account for the clustering history.

In this letter we have presented a semi-numerical approach for the computation of the NNLO quark
jet function for $k_T$-like variables in $e^+e^-$ collisions.
Our method can be applied to any observable in the class (\ref{eq:class}).
The jet function is obtained by integrating collinear splitting kernels over the collinear phase space, and, at NNLO, it receives contributions from real and virtual terms. We have outlined the main steps of their NNLO computation in full generality.
The calculation is affected by rapidity divergences, and we have regulated them
by using a time-like auxiliary vector in the definition of collinear momentum fractions.
We have presented explicit results
for a variable that smoothly captures the $n+ 1$ to $n$ jet transition in $e^+e^-$ collisions, both
in the $E$-scheme and WTA scheme. The calculation can be easily adapted to other $k_T$-like variables.
The results we have presented provide a building block for possible resummed computations
for $k_T$ observables and are directly relevant for NNLO calculations of
multijet cross sections at high-energy colliders when these observables are used as slicing variables \cite{inprep1}.
First results in this direction will be presented elsewhere \cite{inprep2}.
Our results may also be relevant in the matching of fixed-order computations to parton-shower simulations.

\noindent {\bf Acknowledgements}. We would like to thank Thomas Becher, Prasanna
Dhani and Pier Monni for helpful discussions and comments on the manuscript.
This work is supported in part by the Swiss National Science Foundation (SNSF)
under contracts 200020$\_$219367 and 200021$\_$219377. The work of L.B. is funded by the European
Union (ERC, grant agreement No. 101044599, JANUS). Views and opinions expressed
are however those of the authors only and do not necessarily reflect those of
the European Union or the European Research Council Executive Agency. Neither
the European Union nor the granting authority can be held responsible for them.

\appendix

\section{Distributional expansion for the observable-dependent contribution}\label{app:distributional}

For the observable-dependent (subtracted) part of the $C_F^2$ contribution to the jet function we need to expand the factor $1/(z_{1,N}\tilde{z}_{2,N})$ (see Eq.~(\ref{eq:z1z2})) in the limit $k^2_{1,\perp},k^2_{23,\perp}\to 0$ at fixed values of $k^2_{1,\perp}/k^2_{23,\perp}$. In order to do so, we can use the distributional expansion
\begin{equation}
  \label{eq:z1Nz2NDistribution}
  \begin{aligned}
    &\int_0^1 dz_1 \int_0^1 dz_2 \frac{z_1}{z_1^2+y^2 \lambda^2}\frac{z_2}{z_2^2+ \frac{\lambda^2}{(1-z_1)^2}}p(z_1,z_2)=\\
    &\int_0^1 dz_1 \int_0^1 dz_2 \frac{1}{z_1 z_2}\left( p(z_1,z_2)-p_s\left(\frac{z_1}{z_2}\right)-p_{s_1}(z_2)-p_{s_2}(z_1) +p_{s_1,s_2}+p_{s_2,s_1}\right)\\
    &+\int_0^1 \frac{dz_1}{z_1}\log \left(\frac{1-z_1}{\lambda }\right) \left(p_{s_2}(z_1)-p_{s_2,s_1}\right)-\log (\lambda  y)\int_0^1 \frac{dz_2}{z_2} \left(p_{s_1}(z_2)-p_{s_1,s_2}\right)\\
    & +\int_0^1 \frac{dt}{t}\Biggl(\frac{y^2 \log \left(\frac{y}{t}\right) \left(p_s(t)-p_{s_1,s_2}\right)}{t^2-y^2}+\frac{\left(\log \left(\frac{1}{t}\right)-t^2 y^2 \log (y)\right) \left(p_s\left(\frac{1}{t}\right)-p_{s_2,s_1}\right)}{t^2 y^2-1}\\
    &-\log (\lambda ) \left(p_s\left(\frac{1}{t}\right)+p_s(t)-p_{s_1,s_2}-p_{s_2,s_1}\right)\Biggr)+\frac{1}{4} \text{Li}_2\left(1-\frac{1}{y^2}\right) \left(p_{s_1,s_2}-p_{s_2,s_1}\right)\\
    &-\frac{1}{2} \log (\lambda  y) \left(\log (y) \left(p_{s_2,s_1}-p_{s_1,s_2}\right)-\log (\lambda ) \left(p_{s_1,s_2}+p_{s_2,s_1}\right)\right)-\frac{1}{6} \pi ^2 p_{s_2,s_1}+\mathcal{O}(\lambda)\, ,
  \end{aligned}
\end{equation}
where $p(z_1,z_2)$ is an integrable function with well-defined limits
\begin{equation}
  \begin{aligned}
    p_{s_1}(z_2)&=\lim_{z_1\to 0}p(z_1,z_2)\quad\quad p_{s_2}(z_1)=\lim_{z_2\to 0}p(z_1,z_2)\quad\quad p_{s}(t)=\lim_{\lambda\to 0}p(\lambda t,\lambda)\\
    p_{s_1,s_2}&=\lim_{z_2\to 0}\lim_{z_1\to 0}p(z_1,z_2)=\lim_{t\to 0}p_{s}(t)\quad\quad  p_{s_2,s_1}=\lim_{z_1\to 0}\lim_{z_2\to 0}p(z_1,z_2)=\lim_{t\to 0}p_{s}(\frac{1}{t})\, .
  \end{aligned}
\end{equation}

\bibliography{biblio}

\end{document}